%% file: main.tex
\documentclass[sigconf]{acmart}
\usepackage{multirow}
\usepackage{booktabs}
\usepackage{tikz}
\usepackage{pgfplots}
\pgfplotsset{compat=1.18} % Use the latest version for better compatibility
\usetikzlibrary{shapes, arrows, positioning}
\usepackage{fvextra}
\usepackage{tikz}
\usetikzlibrary{arrows.meta, positioning, shapes.geometric, bending}
\usepackage{pgfplots}
\pgfplotsset{compat=1.18}
\usepackage{booktabs}
\AtBeginDocument{%
  }

\setcopyright{rightsretained}
\copyrightyear{2026}
\acmYear{2026}
\acmDOI{}

\acmConference[ACM CAIS 2026]{ACM Workshop on Agentic Software Engineering (ACM CAIS 2026)}{May 26--29, 2026}{San Jose, CA}
\acmISBN{}

\begin{document}

\title{Runtime-Structured Task Decomposition for Agentic Coding Systems}


\author{Shubhi Asthana}
\affiliation{%
  \institution{IBM Research}
  \city{San Jose}
  \country{California}}
\email{sasthan@us.ibm.com}

\author{Bing Zhang}
\authornote{Work done while at IBM Research.}
\affiliation{%
  \institution{Zoom}
  \city{San Jose}
  \country{California}}
\email{bing.zhang@zoom.com}

\author{Chad DeLuca}
\affiliation{%
  \institution{IBM Research}
  \city{San Jose}
  \country{California}}
\email{delucac@us.ibm.com}

\author{Hima Patel}
\affiliation{%
  \institution{IBM Research}
  \city{Bengaluru}
  \country{India}}
\email{himapatel@in.ibm.com}

\author{Ruchi Mahindru}
\affiliation{%
  \institution{IBM Research}
  \city{Yorktown}
  \country{New York}}
\email{rmahindr@us.ibm.com}

\renewcommand{\shortauthors}{Asthana et al.}

\begin{abstract}
Agentic coding systems increasingly deploy LLMs for multi-step software
engineering tasks---debugging, root cause analysis, code review---yet most
encode all task logic, control flow, and output generation within monolithic
prompts. This leads to brittle behavior, poor debuggability, and expensive
full-pipeline reruns when any step fails.
 
We present \emph{runtime-structured task decomposition}, an architectural
pattern in which task partitioning decisions are governed by executable
control flow rather than static prompt text, and LLMs are invoked only for
narrowly scoped judgment tasks with schema-validated outputs.
 
We evaluate this pattern on two software engineering workloads across
\textbf{three configurations}---monolithic, static decomposition (same
subtask graph, no runtime branching), and runtime-structured---over 10 runs
each. Across both workloads, we find that decomposition structure alone does
not reliably reduce retry cost. In the Kubernetes root cause analysis
workload, the static baseline's retry cost ($1{,}632 \pm 145$ tokens)
exceeds the monolithic baseline ($904 \pm 17$ tokens) by 80.5\%,
because fixed sequential execution must rerun multiple downstream
subtasks. In the multi-file debugging workload, the same effect
appears at smaller magnitude (933 vs.\ 703 tokens). Runtime-structured
decomposition, by retrying only the failed subtask, reduces retry cost
in both settings---to $436 \pm 132$ tokens (RCA) and 460 tokens
(debugging)---achieving up to a $51.7\%$ reduction over monolithic
and a $73.2\%$ reduction over static decomposition.
 We discuss implications for agentic coding system design,
production debugging, and AgentOps workflows.
\end{abstract}





\maketitle

\input{introduction} 
\input{system_overview}
\input{demo}
\input{related_work}
\input{conclusion}

\bibliographystyle{plainnat}
\bibliography{output}

\end{document}

%% file: introduction.tex
\section{Introduction}
\label{sec:Introduction}

Agentic coding systems---such as those built on AutoGen, LangGraph, and
Claude Code---are among the earliest LLM agents deployed in real software
engineering workflows. When a debugging agent fails, does it retry the full
analysis or just the broken step? When a root cause analysis pipeline
produces a malformed output, can it recover without re-ingesting thousands
of log lines? These are production engineering questions with architectural
answers.

%Most existing agentic coding systems encode task logic, control flow, and output generation within monolithic prompts. While effective for simple workflows, this design struggles as tasks grow in complexity:

\iffalse
\begin{enumerate}
  \item \textbf{Brittleness:} Small changes in input structure or prompt
    wording can disrupt the entire workflow.
  \item \textbf{Limited debuggability:} When reasoning and orchestration are
    entangled within a single prompt, failures are difficult to isolate.
  \item \textbf{Expensive recovery:} Any failure requires a full pipeline
    rerun, paying the cost of re-ingesting all prior context.
\end{enumerate}
\fi

Most existing agentic coding systems encode task logic, control
flow, and output generation within monolithic prompts, leading to
brittle behavior, limited debuggability, and expensive full-pipeline
reruns when failures occur.

Recent frameworks such as DSPy~\cite{dspy2024}, LangGraph~\cite{langgraph2024},
and AutoGen~\cite{wu2023autogen} have introduced programmatic abstractions
for orchestrating LLM calls. However, task decomposition strategies in these
frameworks are typically specified statically by the developer---the structure
of the task must be known and fixed at design time. When intermediate outputs
are invalid or branching conditions emerge at runtime, a static decomposition
cannot adapt.

In this paper, we present \textbf{runtime-structured task decomposition (RSTD)}, an
architectural pattern in which task partitioning decisions are governed by
executable control flow and validated intermediate signals rather than static
prompt text. Under this paradigm:
\begin{enumerate}
\item LLMs are invoked only for narrowly scoped judgment tasks with
  schema-validated outputs.
\item Orchestration, branching, and state management are handled
  deterministically in code.
\item Failures are isolated to individual subtasks, enabling selective retry
  without full pipeline rerun.
\end{enumerate}

We implement this pattern using the Mellea generative computing
framework~\cite{mellea2025} and evaluate it across \textbf{three
configurations}---monolithic, static decomposition (same subtask structure,
no runtime branching), and runtime-structured---over 10 runs each. The
three-way comparison isolates the contribution of runtime branching from
decomposition structure alone.
 
A key empirical finding: static decomposition alone does not
reliably reduce retry cost relative to monolithic, and can
increase it when failures trigger cascading re-execution of
downstream subtasks.
Its retry cost ($1{,}632 \pm 145$ tokens)
\emph{exceeds} the monolithic baseline ($904 \pm 17$ tokens) by 80.5\%,
because fixed sequential execution must rerun all downstream subtasks from
the point of failure. Runtime-structured decomposition, by retrying only
the failed subtask ($436 \pm 132$ tokens), achieves a $51.7\%$ retry cost
reduction over monolithic and a $73.2\%$ reduction over static
decomposition---demonstrating that runtime branching, rather than decomposition
structure alone, is the primary mechanism.

Our contributions are:
\begin{enumerate}
  \item A formalization of RSTD as an architectural pattern for agentic
    coding systems, distinct from both monolithic prompting and static
    decomposition in DSPy, LangGraph, and AutoGen.
  \item A three-configuration empirical evaluation across 10 runs per
    setting, with variance quantified, framework overhead separated from
    LLM API latency, and a static baseline that isolates the contribution
    of runtime branching.
  \item Operational implications for agentic coding systems: subtask-level
    monitoring, selective retry under failure, and per-subtask model
    substitution.
\end{enumerate}

%% file: system_overview.tex
\section{Architecture}
\label{sec:system_overview}

Runtime-structured task decomposition separates orchestration, state
management, and LLM inference into distinct layers. Rather than embedding
task structure inside prompts, decomposition decisions are externalized into
executable control flow, enabling task structure to be determined dynamically
at runtime.

The system consists of three components, illustrated in
Figure~\ref{fig:adaptive_architecture}:

\begin{figure}[t]
    \centering
    \includegraphics[width=0.9\linewidth]{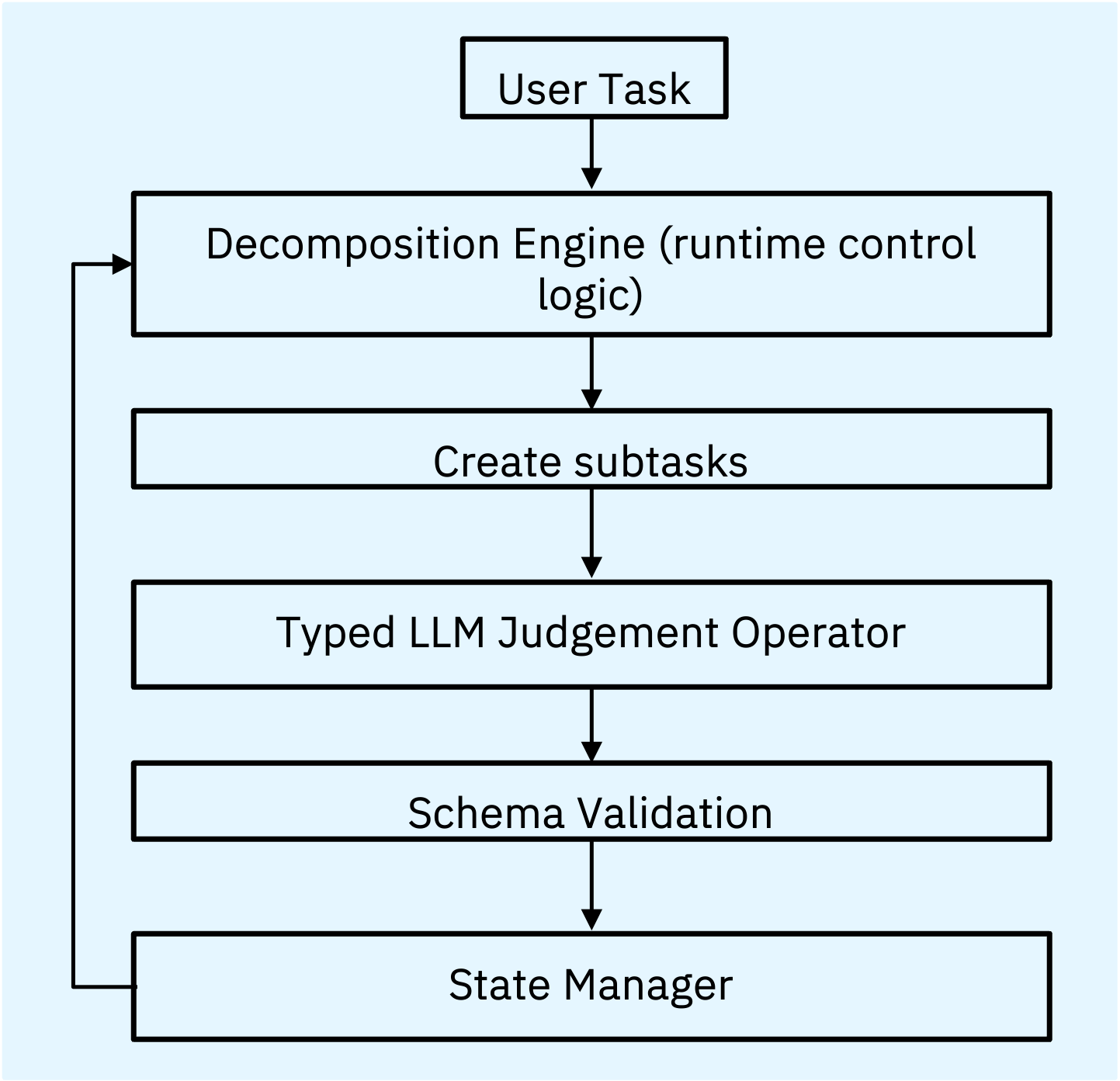}
    \caption{Runtime-structured decomposition architecture. A decomposition
    engine governs control flow, invoking typed LLM judgment operators over
    bounded context. Outputs are validated and stored in persistent state,
    feeding back into subsequent decomposition decisions.}
    \label{fig:adaptive_architecture}
\end{figure}

\begin{itemize}
\item \textbf{Decomposition Engine}: Governs runtime task partitioning via
  developer-authored conditional logic---branching on schema validation
  outcomes, confidence thresholds, or predecessor completion status.
  Branching decisions are resolved at runtime against validated intermediate
  state.

\item \textbf{Judgment Operators}: Typed LLM calls scoped to a single
  reasoning task. Each call declares an explicit output schema; validation
  failure triggers a targeted repair prompt rather than full pipeline rerun.

\item \textbf{State Manager}: Persists validated intermediate results keyed
  by subtask identifier. Downstream subtasks access prior state explicitly
  by key, keeping per-call context bounded.
\end{itemize}

Branching decisions evaluate three runtime signals: (1)~schema validation
outcome; (2)~output content (non-empty result or confidence above threshold);
and (3)~subtask completion status. All policies are deterministic conditional
logic, making them auditable, unit-testable, and reproducible.

\begin{figure}[t]
\centering
\begin{tikzpicture}[
  box/.style={draw, rounded corners, minimum width=1.0cm, minimum height=0.5cm,
              font=\scriptsize, align=center},
  failbox/.style={box, fill=red!20},
  retrybox/.style={box, fill=orange!30},
  lbl/.style={font=\scriptsize\bfseries, anchor=east},
  arr/.style={-{Stealth}, thick}
]

%% Row 1: Monolithic  (y=2.2)
\node[lbl] at (0.55,2.2) {Mono.};
\node[box, minimum width=3.2cm] (mono) at (2.8,2.2) {Monolithic Prompt\\(all logic)};
\node[failbox] (mfail) at (5.0,2.2) {Fail};
\node[retrybox, minimum width=3.2cm] (mretry) at (2.8,1.6) {Full Rerun};
\draw[arr] (mono) -- (mfail);
\draw[arr, dashed] (mfail) -- ++(0,-0.32) -| (mretry);

%% Row 2: Static  (y=0.8)
\node[lbl] at (0.55,0.8) {Static};
\foreach \i/\x in {1/1.0, 2/2.2, 3/3.4, 4/4.6} {
  \node[box] (s\i) at (\x,0.8) {S\i};
}
\node[failbox] (sfail) at (5.7,0.8) {Fail};
\draw[arr] (s1)--(s2); \draw[arr] (s2)--(s3);
\draw[arr] (s3)--(s4); \draw[arr] (s4)--(sfail);
\draw[arr, dashed] (sfail) -- ++(0,-0.32) -| (s2);
\node[font=\tiny, text=red, above] at (3.4,0.45) {reruns S2\,S3\,S4};

%% Row 3: RSTD  (y=-0.5)
\node[lbl] at (0.55,-0.5) {RSTD};
\foreach \i/\x in {1/1.0, 2/2.2, 3/3.4, 4/4.6} {
  \node[box] (r\i) at (\x,-0.5) {S\i};
}
\node[failbox] (rfail) at (5.7,-0.5) {Fail};
\node[retrybox] (rretry) at (3.4,-1.2) {Retry S3 only};
\draw[arr] (r1)--(r2); \draw[arr] (r2)--(r3);
\draw[arr] (r3)--(r4); \draw[arr] (r4)--(rfail);
\draw[arr, dashed] (rfail) -- ++(0,-0.38) -- (rretry.east);
\draw[arr] (rretry.north) -- (r3.south);

\end{tikzpicture}
\caption{Retry behavior under subtask failure. Monolithic reruns the full
pipeline; static reruns all downstream subtasks; RSTD retries only the failed
subtask.}
\label{fig:retry_comparison}
\end{figure}
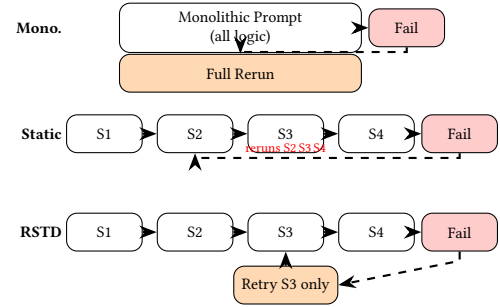

We operationalize this architecture using the open-source \textbf{Mellea
framework}~\cite{mellea2025}, which provides typed LLM calls with enforced
input/output schemas, persistent session state, and automated
validation-and-repair loops. A Mellea call specifies a prompt and a set of
output requirements:

\begin{verbatim}
result = session.instruct(
  prompt = "Classify anomalies. Output JSON: "
           "[{anomaly, type, severity, conf}]",
  requirements = [
    "Output must be a JSON list.",
    "Each item must have: anomaly, type,
     severity, confidence."
  ]
)
# On failure, Mellea retries with the
# validation error appended.
\end{verbatim}

%% file: demo.tex
\section{Case Studies}
\label{sec:demo}

We demonstrate RSTD through two software engineering workloads evaluated
across \textbf{three configurations}: (i)~\textbf{Monolithic}---all task
logic in a single prompt; (ii)~\textbf{Static}---same subtask graph, fixed
sequential execution, no runtime branching; and (iii)~\textbf{RSTD}---our
approach, with runtime-controlled branching and selective retry. The static
baseline isolates whether retry advantages come from decomposition structure
alone or specifically from runtime branching.

All agents use \texttt{gpt-4} at temperature~0. Token counts are measured
using \texttt{tiktoken}. Each configuration is run \textbf{10 times}; we
report mean $\pm$ standard deviation. Non-zero standard deviations at
temperature~0 reflect API-side variability (e.g., request scheduling and
batch-size variation) rather than sampling randomness. Monolithic baselines
are drawn from \url{https://github.com/microsoft/autogen} and
\url{https://github.com/fuzzylabs/sre-agent}; structured agent code uses
\url{https://mellea.ai/}.

\subsection{Schema Validation Failure Rate}
\label{subsec:failure_rate}

Natural schema validation failure rates were low: 0\% across 40 subtask
executions in UC1, and 2.0\% (2/100) in UC2, both at Subtask~2
(Anomaly Classification) due to a missing \texttt{confidence} field.
Retry cost is therefore measured under a \emph{simulated} failure---a
structurally malformed upstream input (one missing JSON field)---which
measures recovery cost as an architectural property rather than a
frequency estimate.

\subsection{Use Case 1: Multi-File Code Debugging}
\label{subsec:uc1}
 
The \textbf{monolithic} baseline uses AutoGen's
\texttt{AssistantAgent}~\cite{wu2023autogen}, encoding bug identification,
fix generation, validation, and report synthesis within a single system
message.
 
The task involves a two-file Python pipeline with three causally ordered
bugs: (1)~a wrong aggregation operator in \texttt{pipeline.py};
(2)~an off-by-one sliding window error masked by bug~1; (3)~a type mismatch
in \texttt{validator.py} unreachable until bugs~1 and~2 are fixed.
 
The \textbf{static} baseline uses the same four subtasks---Analysis, Fix
Generation, Validation, Synthesis---in a fixed sequence. On Validation failure, it reruns Fix Generation and Validation,
introducing limited cascading re-execution.
 
The \textbf{runtime-structured} agent uses the same four subtasks but retries
only Fix Generation on Validation failure, consuming only the bug analysis
JSON---not the full source files.
 
All three configurations correctly identified and fixed all three bugs across
all 10 runs (100\% correctness). Results are in Table~\ref{tab:uc1}.
Static decomposition increases retry cost relative to monolithic
(933 vs.\ 703 tokens, +32.7\%), though more modestly than in Use Case~2
due to the shallower dependency chain (two downstream subtasks vs.\ three).
RSTD avoids cascading re-execution by retrying only the failed subtask,
reducing retry cost to 460 tokens (34.6\% below monolithic).

Note that RSTD's higher baseline token cost (2,225 vs.\ 703 tokens)
reflects the overhead of making multiple API calls for normal successful
execution. This overhead is worthwhile when failure rates are non-trivial
or when subtasks involve expensive context re-ingestion---the primary
deployment scenario this pattern targets.

\begin{table}[h]
\caption{Use Case 1: Multi-file code debugging across 10 runs. Retry cost
measured under simulated Validation failure
(Section~\ref{subsec:failure_rate}).}
\label{tab:uc1}
\begin{tabular}{lrrr}
\toprule
\textbf{Metric} & \textbf{Mono.} & \textbf{Static} & \textbf{RSTD} \\
\midrule
Tokens (mean $\pm$ sd)
  & $703 \pm 49$
  & $2181 \pm 240$
  & $2225 \pm 270$ \\
Latency s (mean $\pm$ sd)
  & $15.37 \pm 18.38$
  & $26.28 \pm 2.59$
  & $21.91 \pm 2.70$ \\
\quad LLM API (s)
  & $15.37 \pm 18.38$
  & $21.55 \pm 2.13$
  & $17.96 \pm 2.21$ \\
\quad Framework (s)
  & ---
  & $4.73 \pm 0.47$
  & $3.94 \pm 0.49$ \\
LLM calls
  & 1
  & 4
  & 4 \\
Correct (all runs)
  & 100\%
  & 100\%
  & 100\% \\
Retry tokens
  & $703 \pm 49$
  & $933 \pm 93$
  & $460 \pm 77$ \\
\bottomrule
\end{tabular}
\end{table}

\subsection{Use Case 2: Kubernetes Root Cause Analysis}
\label{subsec:uc2}
 
The \textbf{monolithic} baseline replicates
\texttt{fuzzylabs/sre-agent}~\cite{fuzzylabs_sre}, encoding log parsing,
anomaly detection, root cause inference, remediation planning, and report
generation within a single prompt.
 
The task involves a Kubernetes OOMKill on \texttt{payment-svc} following
deployment v2.4.2, with database connection pool exhaustion and a p99 latency
SLA breach. Ground truth: a deployment regression introducing unbounded
in-memory batch processing of 84,000 invoice records.
 
The \textbf{static} baseline uses the same five subtasks in a fixed sequence.
On Subtask~3 (RCA) failure, it reruns Subtasks~3, 4, and 5.
 
\begin{figure*}[t]
  \centering
  \includegraphics[width=\linewidth]{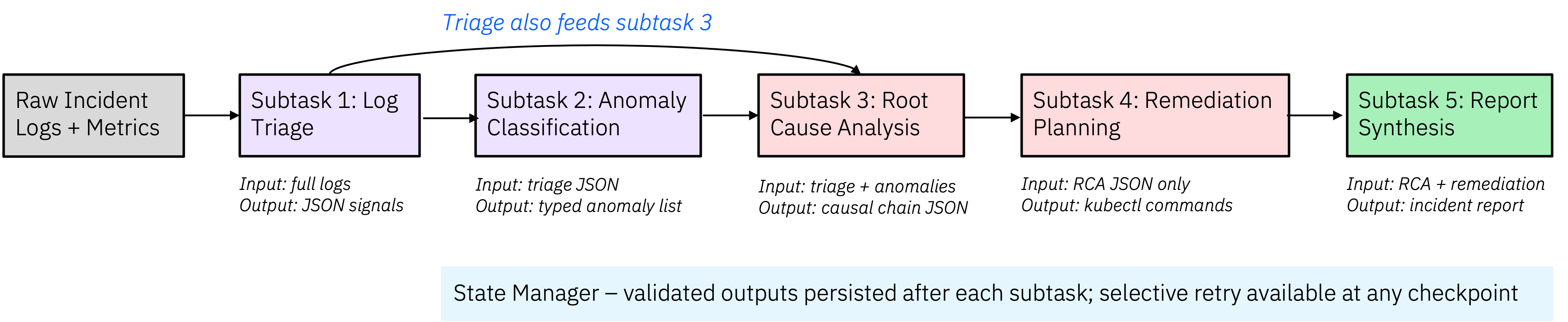}
  \caption{Runtime-structured SRE pipeline. Each subtask receives only
  predecessor outputs. Triage feeds both Subtask~2 and Subtask~3 (skip arc).
  Outputs are schema-validated before State Manager writes.}
  \label{fig:sre_pipeline}
\end{figure*}

The \textbf{runtime-structured} agent decomposes into five Mellea subtasks,
each receiving only the context it requires (Figure~\ref{fig:sre_pipeline}).
On RCA failure, only Subtask~3 is retried.
 
All configurations correctly identified the root cause across all 10 runs.
Results are in Table~\ref{tab:uc2}.
 
\begin{table}[h]
\caption{Use Case 2: Kubernetes root cause analysis across 10 runs. Retry
cost measured under simulated RCA failure at Subtask~3
(Section~\ref{subsec:failure_rate}).}
\label{tab:uc2}
\begin{tabular}{lrrr}
\toprule
\textbf{Metric} & \textbf{Mono.} & \textbf{Static} & \textbf{RSTD} \\
\midrule
Tokens (mean $\pm$ sd)
  & $904 \pm 17$
  & $2{,}553 \pm 224$
  & $2{,}716 \pm 424$ \\
Latency s (mean $\pm$ sd)
  & $10.41 \pm 2.31$
  & $22.76 \pm 2.52$
  & $28.78 \pm 5.06$ \\
\quad LLM API (s)
  & $10.41$ 
  & $18.66 \pm 2.07$
  & $23.60 \pm 4.15$ \\
\quad Framework (s)
  & ---
  & $4.10 \pm 0.45$
  & $5.18 \pm 0.91$ \\
LLM calls          & 1     & 5     & 5     \\
Correct (all runs) & 100\% & 100\% & 100\% \\
Retry tokens (mean $\pm$ sd)
  & $904 \pm 17$
  & $1{,}632 \pm 145$
  & $436 \pm 132$ \\
\bottomrule
\end{tabular}
\end{table}
 
The key finding from Table~\ref{tab:uc2} is counterintuitive: the static
baseline's retry cost ($1{,}632 \pm 145$ tokens) \emph{exceeds} the
monolithic baseline ($904 \pm 17$ tokens) by 80.5\%, because fixed
sequential execution must rerun Subtasks~3, 4, and~5 on failure.
RSTD, by retrying only Subtask~3 ($436 \pm 132$ tokens), achieves a
\textbf{51.7\%} retry cost reduction over monolithic and a
\textbf{73.2\%} reduction over static decomposition---confirming that
\emph{runtime branching}, not decomposition structure alone, is the
primary recovery cost mechanism.

Framework overhead accounts for $4.10$--$5.18$s ($\sim$18\% of total
latency); the latency gap between monolithic ($10.41$s) and decomposed
agents ($\sim$23--29s) reflects additional API round-trips, not
orchestration inefficiency.
 
\subsection{Design Implications}
\label{subsec:implications}
 
\textit{Selective retry reduces recovery cost under failure.}
RSTD achieves a 73.2\% retry cost reduction over static and 51.7\% over
monolithic. Static decomposition without runtime branching can \emph{increase}
retry cost relative to monolithic when pipeline depth causes cascading
reruns---in UC2, by 80.5\%. The higher RSTD baseline cost is a fixed overhead;
whether it is offset by retry savings depends on deployment failure rates.

\textit{Subtask-level monitoring enables targeted AgentOps.}
Each subtask boundary is a natural instrumentation point; natural failure data
(Section~\ref{subsec:failure_rate}) identifies Subtask~2 as the sole natural
failure point across both use cases.

\textit{Per-subtask model substitution is structurally available.}
Framework overhead is consistent ($\sim$18\% of wall time), so subtasks can
be assigned different models without reconfiguring the orchestration layer.
 

%% file: related_work.tex
\section{Related Work}
\label{sec:related_work}

\textbf{Agentic coding systems.} SWE-bench~\cite{swebench2024} and
AgentBench~\cite{agentbench2024} evaluate agent capability on
repository-level tasks. Our work addresses the complementary
\emph{operational} gap: not whether agents can solve coding tasks, but
how to architect them for reliable, debuggable, cost-efficient operation.

\textbf{Programmatic LLM frameworks.} DSPy~\cite{dspy2024} proposes
declarative LLM programs with typed modules and prompt optimization.
LangGraph~\cite{langgraph2024} provides graph-based orchestration for
multi-step pipelines. AutoGen~\cite{wu2023autogen} enables multi-agent
conversation patterns. These frameworks support static decomposition---the
task graph is fixed at design time. Our pattern differs in that branching
and recovery decisions are resolved at runtime against validated intermediate
state, and validation failure blocks downstream state transitions entirely
rather than passing malformed outputs forward.

The closest related mechanism is DSPy Assertions~\cite{singhvi2024dspy},
which enforce semantic constraints within modules. The key distinction: DSPy
Assertions operate within a module and trigger prompt optimization as the
primary recovery path. Our validation-gated state transitions operate
\emph{between} subtasks---a failed subtask's output is not written to the
State Manager and is never visible to downstream subtasks, preventing
cascading reasoning errors.

\textbf{Inference-time control.} Reflexion~\cite{reflexion} proposes verbal
reinforcement for iterative self-correction. Its retry mechanism operates at
the level of the full task response. Our validation-gated recovery retries
only the failed subtask over its bounded context, providing finer-grained
and more token-efficient error correction. LLMCompiler~\cite{llmcompiler}
parallelizes a statically derived task graph; our approach adapts branching
at runtime without a pre-compiled execution plan.

\textbf{Production operations for LLM systems.} ITBench~\cite{itbench2024}
evaluates agents on IT automation tasks including incident response.
Our SRE use case addresses the same domain; future evaluation on ITBench
will quantify whether structured decomposition improves accuracy on diverse
incident types at scale.

%% file: conclusion.tex
\section{Conclusion}
\label{sec:conclusion}

We presented RSTD as an architectural pattern that externalizes task structure
into executable control flow, enabling selective retry at subtask granularity.
Key findings across three configurations over 10 runs:
(1)~static decomposition does not reliably reduce retry cost---its cost
($1{,}632 \pm 145$ tokens) exceeds monolithic ($904 \pm 17$ tokens) by
80.5\% due to cascading re-execution;
(2)~RSTD achieves a \textbf{51.7\%} retry cost reduction over monolithic
and \textbf{73.2\%} over static ($436 \pm 132$ tokens);
and (3)~framework overhead accounts for $\sim$18\% of total latency.

\textit{Limitations.}
Both use cases are controlled scenarios at temperature~0 with low natural
failure rates (0--2\%), so retry cost is measured under simulated failure.
The higher RSTD baseline cost means overall token savings depend on
deployment failure rates, which we do not measure at scale.
Decomposition policies are developer-authored and may not generalize to
automatically derived graphs.

Future work includes evaluation on ITBench~\cite{itbench2024} and broader
benchmarks, extending to multi-agent settings, and learning decomposition
policies from runtime signals.

%Three directions remain for future work. First, evaluation on ITBench~\cite{itbench2024} and broader software engineering benchmarks will quantify accuracy and cost trade-offs across diverse tasks at scale. Second, extending structured decomposition to multi-agent coding systems---where subtask outputs are shared across agents---is a natural next step. Third, current decomposition strategies are developer-authored; learned decomposition policies that automatically partition tasks based on runtime signals would reduce the engineering burden of deploying this pattern.

%% file: output.bib
@misc{singhvi2024dspy,
  title={{DSPy} Assertions: Computational Constraints for
         Self-Refining Language Model Pipelines},
  author={Arnav Singhvi and Manish Shetty and Shangyin Tan and
          Christopher Potts and Koushik Sen and Matei Zaharia
          and Omar Khattab},
  year={2024},
  eprint={2312.13382},
  archivePrefix={arXiv},
  primaryClass={cs.CL}
}

@article{wu2023autogen,
  title={AutoGen: Enabling Next-Gen LLM Applications via Multi-Agent Conversation},
  author={Wu, Qingyun and Bansal, Gagan and Zhang, Jie and Wu, Yiran and Li, Bei and Zhu, Erkang and Jiang, Li and Zhang, Xiaoyun and Liu, Zhiyuan and others},
  journal={arXiv preprint arXiv:2308.08155},
  year={2023}
}

@misc{langgraph2024,
  title={LangGraph},
  author={{LangChain AI}},
  year={2024},
  howpublished={\url{https://github.com/langchain-ai/langgraph}}
}

@article{swebench2024,
  title={SWE-bench: Can Language Models Resolve Real-World GitHub Issues?},
  author={Jimenez, Carlos E. and Yang, John and Wettig, Alexander and others},
  journal={arXiv preprint arXiv:2310.06770},
  year={2024}
}

@article{llmcompiler,
  title   = {{An LLM Compiler for Parallel Function Calling}},
  author  = {Kim, Sehoon and Moon, Suhong and Tabrizi, Rohan and
             Lee, Nicholas and Mahoney, Michael and Keutzer, Kurt
             and Gholami, Amir},
  journal = {arXiv preprint arXiv:2312.04511},
  year    = {2023}
}

@inproceedings{reflexion,
  title     = {{Reflexion: Language Agents with Verbal Reinforcement Learning}},
  author    = {Shinn, Noah and Cassano, Federico and Gopinath, Ashwin
               and Narasimhan, Karthik and Yao, Shunyu},
  booktitle = {Advances in Neural Information Processing Systems},
  year      = {2023}
}

@article{agentbench2024,
  title={AgentBench: Evaluating LLMs as Agents},
  author={Liu, Xiao and others},
  journal={arXiv preprint arXiv:2308.03688},
  year={2023}
}

@misc{fuzzylabs_sre,
  author       = {{FuzzyLabs}},
  title        = {{SRE Agent: A Site Reliability Engineer AI Agent}},
  year         = {2024},
  howpublished = {\url{https://github.com/fuzzylabs/sre-agent}}
}

@article{itbench2024,
  title={Itbench: Evaluating ai agents across diverse real-world it automation tasks},
  author={Jha, Saurabh and Arora, Rohan and Watanabe, Yuji and Yanagawa, Takumi and Chen, Yinfang and Clark, Jackson and Bhavya, Bhavya and Verma, Mudit and Kumar, Harshit and Kitahara, Hirokuni and others},
  journal={arXiv preprint arXiv:2502.05352},
  year={2025}
}

@article{dspy2024,
  title={DSPy: Programming Language Models Instead of Prompting Them},
  author={Khattab, Omar and others},
  journal={arXiv preprint arXiv:2310.03714},
  year={2024}
}

@misc{mellea2025,
  title={Mellea: A Generative Computing Framework for Structured LLM Programs},
  author={{Mellea Contributors}},
  year={2025},
  howpublished={\url{https://mellea.ai/}}
}
